\documentclass[,final]{aipproc}
\layoutstyle{6x9}
\begin{document}
\title{Highly synchronized noise-driven oscillatory behavior of a FitzHugh--Nagumo ring\\
with phase-repulsive coupling}
\author{Gonzalo Iz\'us}{address={Departamento de F\'{\i}sica, Facultad de Ciencias Exactas y Naturales,\\Universidad Nacional de Mar del Plata,\\De\'an Funes 3350, 7600 Mar del Plata, Argentina.},altaddress={Member, CONICET}}
\author{Roberto Deza}{address={Departamento de F\'{\i}sica, Facultad de Ciencias Exactas y Naturales,\\Universidad Nacional de Mar del Plata,\\De\'an Funes 3350, 7600 Mar del Plata, Argentina.}}
\author{Alejandro S\'anchez}{address={Departamento de F\'{\i}sica, Facultad de Ciencias Exactas y Naturales,\\Universidad Nacional de Mar del Plata,\\De\'an Funes 3350, 7600 Mar del Plata, Argentina.},altaddress={Member, CONICET}}
\begin{abstract}
We investigate a ring of $N$ FitzHugh--Nagumo elements coupled in \emph{phase-repulsive} fashion and submitted to a (subthreshold) common oscillatory signal and independent Gaussian white noises. This system can be regarded as a reduced version of the one studied in [Phys.\ Rev.\ E \textbf{64}, 041912 (2001)], although externally forced and submitted to noise. The noise-sustained synchronization of the system with the external signal is characterized.
\end{abstract}
\maketitle
\section{\label{sec:1}Introduction}
In Ref.\ \cite{bcnm01}---through comparison with the synchronization patterns arising in two-dimensional arrays of FitzHugh--Nagumo (FHN) elements with \emph{phase-repulsive} linear nearest-neighbor coupling---the authors were able to conclude that \emph{intracellular} calcium oscillations in cultures of human epileptic astrocytes \emph{do} interact, since the phases of nearby oscillating astrocytes maintain a nontrivial relationship. It is a fortunate fact that the (space-independent) FHN model is one of the very few multicomponent systems for which a \emph{nonequilibrium potential} (NEP) has been found \cite{izdw98,izdw99}, since NEPs allow in general for a deep insight on the dynamical mechanisms leading to pattern formation and other phenomena where fluctuations play a constructive role \cite{hwio97}. The (albeit minimal) extension of the result in Refs.\ \cite{izdw98,izdw99} towards extended systems carried out in this work is however enough to shed light on the dynamical cause of the conclusion in Ref.\ \cite{bcnm01}: a dynamical symmetry breakdown takes place because the phase-repulsive coupling minimizes the corresponding NEP. When the system is externally forced with a frequency less than the typical inverse deterministic time the cycle duplicates, breaking down into an ``excited'' phase and an ``inhibited'' one. These phases force neighbor elements to alternate with the one in between, thus creating a nontrivial phase relationship between nearby oscillating elements.

\begin{figure}\label{fig:1}
\includegraphics[height=.3\textheight,bb= 0pt 0pt 659pt 401pt]{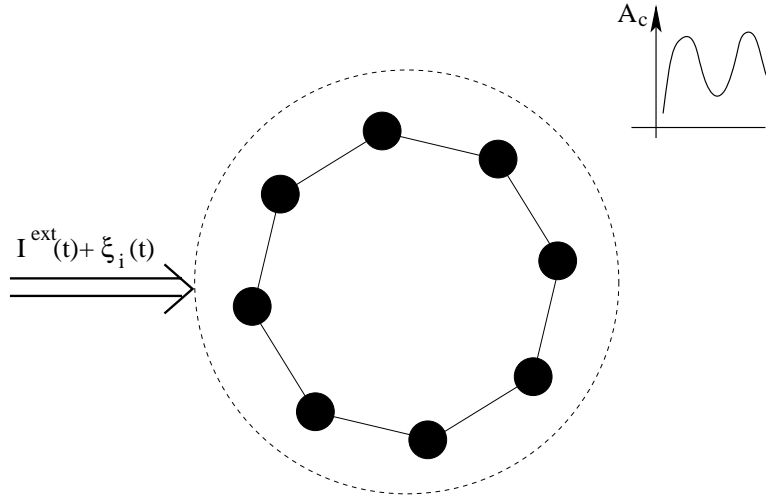}
\caption{Sketch of the system and of its response $A_c(t)$.}
\end{figure}

The system we consider is sketched in Fig.\ \ref{fig:1}: a ring of $N=256$ identical FHN elements with \emph{phase-repulsive} nearest-neighbor coupling and submitted to a (subthreshold) common oscillatory signal and independent Gaussian white noises $\xi_{u_i}(t)$, $\xi_{v_i}(t)$ with $\langle\xi_m(t)\xi_n(t')\rangle=2\eta\delta_{mn}\delta(t-t')$, $m,n=1,\ldots,2N$.
The set of equations governing its dynamics is
\begin{eqnarray}
\dot{u}_i&=&a_c\,u_i\,(1-u_i^2)-v_i+S_g(t)-D(u_{i+1}+u_{i-1})+r_1\,\xi_{u_i}(t)+r_2\,\xi_{v_i}(t)\label{eq:FHN}\\
\dot{v}_i&=&\epsilon\,(\beta\,u_i-v_i+C)+r_3\,\xi_{u_i}(t)+r_4\,\xi_{v_i}(t),\quad i=1,\ldots,N,\quad u_{N+1}=u_1.
\nonumber
\end{eqnarray}
where $\epsilon=0.01$ is the ratio between the relaxation rates of $u_i$ and $v_i$, $\beta=0.01$, $a_c=0.06$ and $C=0.02$ is a suitable constant to set the rest point in Fig.\ \ref{fig:2}a. $D=0.01$ is the \emph{phase-repulsive} coupling constant, and the $r_i$ (which determine the transport matrix) are $r_1=0.998\times10^2$, $r_2=0.499\times10^1$, $r_3=0.998$, $r_4=0.499\times10^{-1}$. Moreover, taking the Milshtein integration step as $dt=5\times10^{-3}$, we estimate the typical inverse deterministic time as $0.838\times10^{-3}$ and so we take the excitation frequency $\Omega_0$ as a fraction of that value (typically 0.1--0.4). Given that, $S_g(t)=0.0275\sin\Omega_0t$.
\section{\label{sec:2}The nonequilibrium potential}
\begin{figure}\label{fig:2}
  \includegraphics[height=.2\textheight,bb= 54pt 360pt 558pt 720pt]{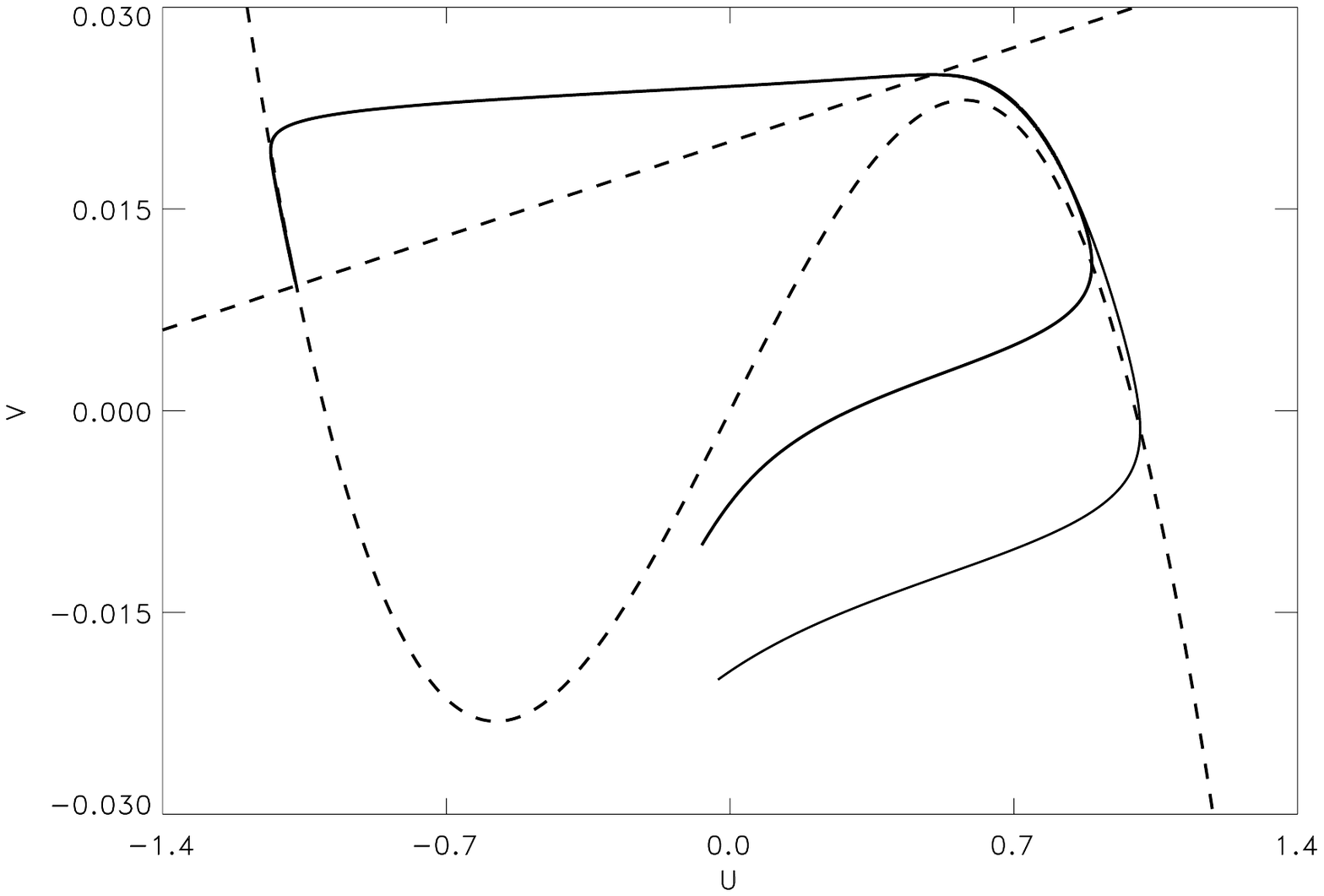}
  \includegraphics[height=.2\textheight,bb= 54pt 360pt 558pt 720pt]{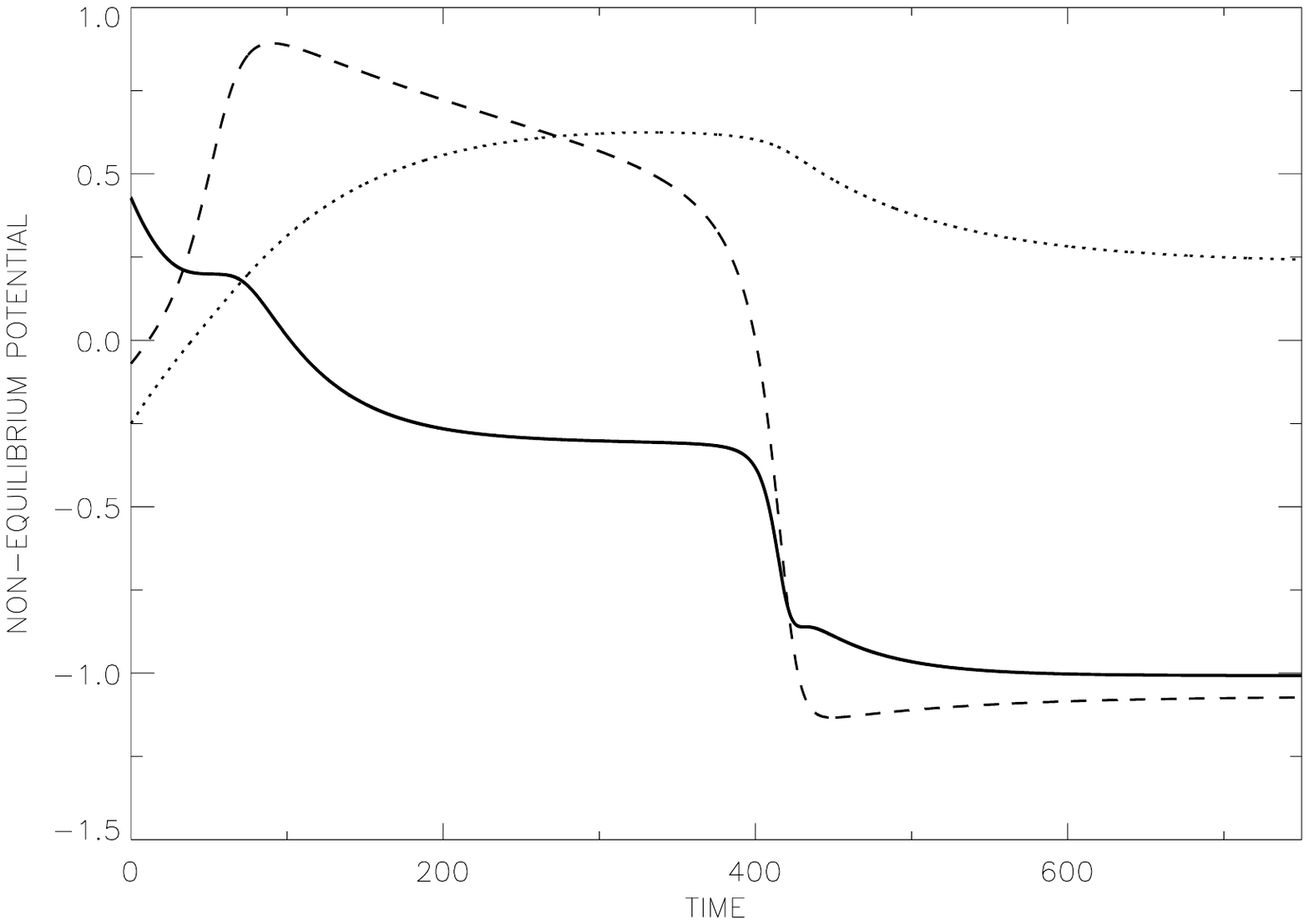}
  \caption{a) Phase-space excursions in excitable regime (the nullclines are indicated in dashed line); b) Time evolution of $u$ (dashed line), $v$ (dotted line), and the NEP (full line) during a phase-space excursion. The scales of $v$ and the NEP were adjusted for better comparison.}
\end{figure}
\begin{figure}\label{fig:3}
  \includegraphics[height=.2\textheight,bb= 54pt 360pt 558pt 720pt]{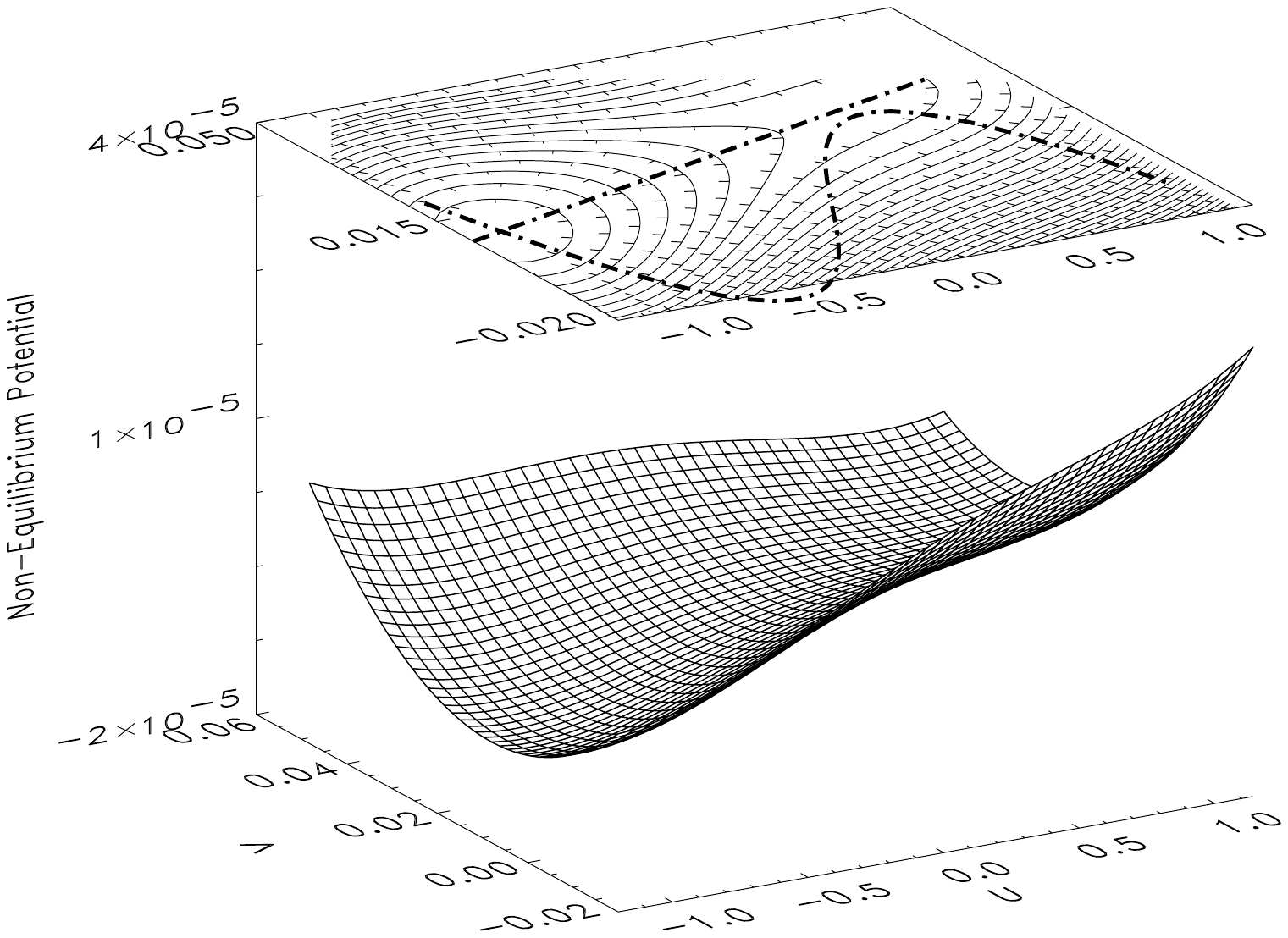}
  \includegraphics[height=.2\textheight,bb= 54pt 360pt 558pt 720pt]{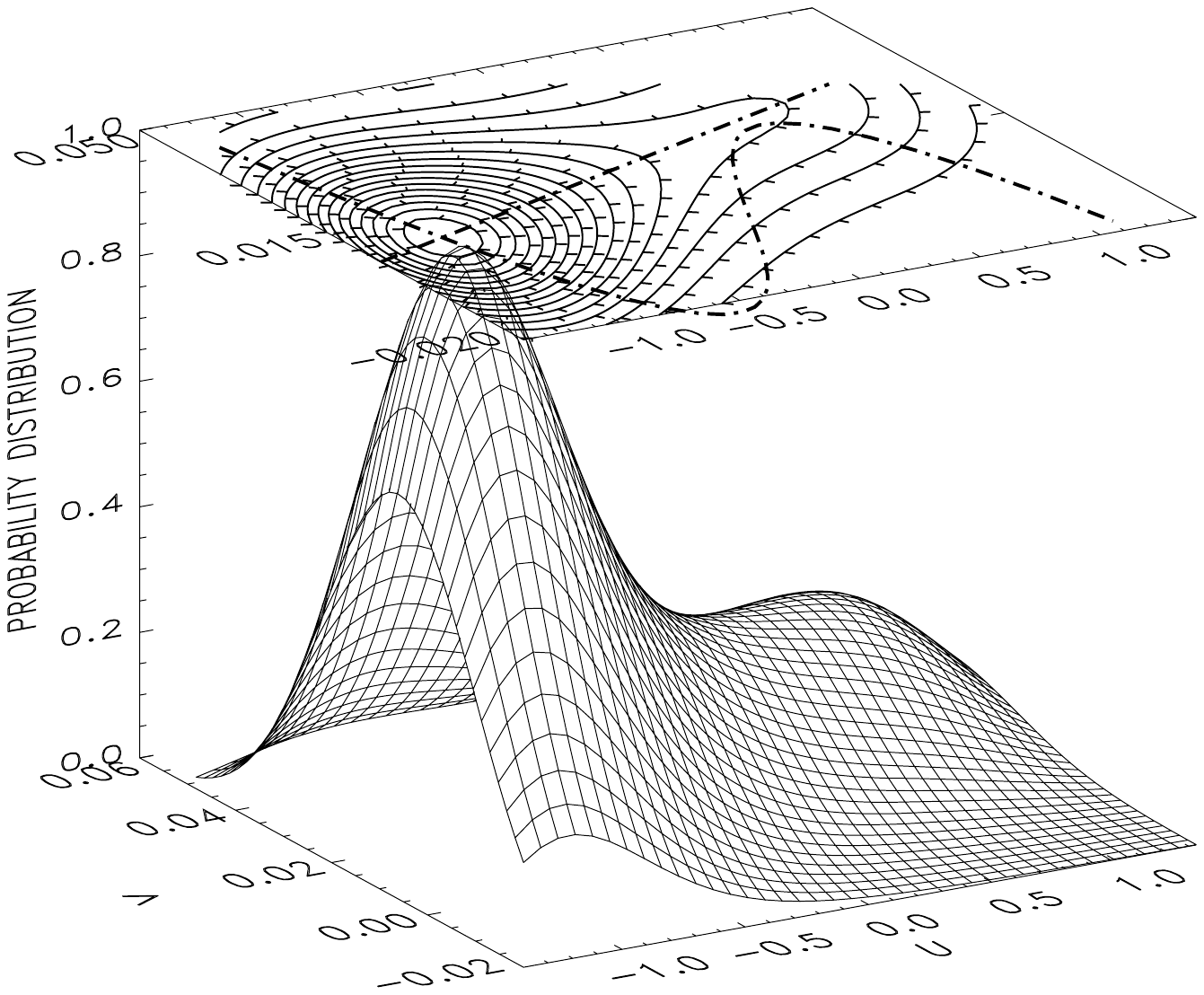}
  \caption{a) NEP in excitable regime; b) Stationary pdf in excitable regime.}
\end{figure}
\begin{figure}\label{fig:4}
  \includegraphics[height=.2\textheight,bb= 54pt 360pt 558pt 720pt]{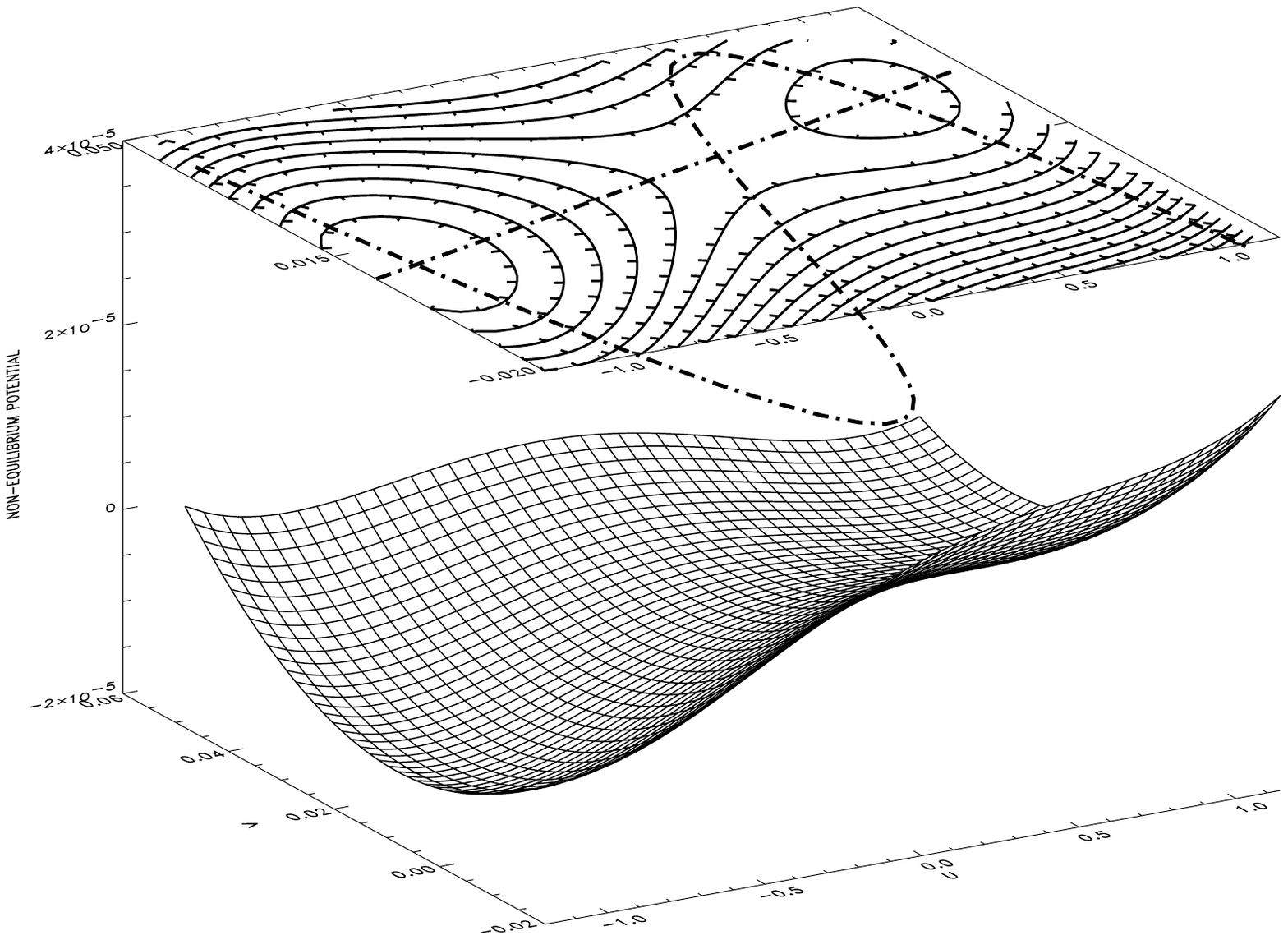}
  \includegraphics[height=.2\textheight,bb= 54pt 360pt 558pt 720pt]{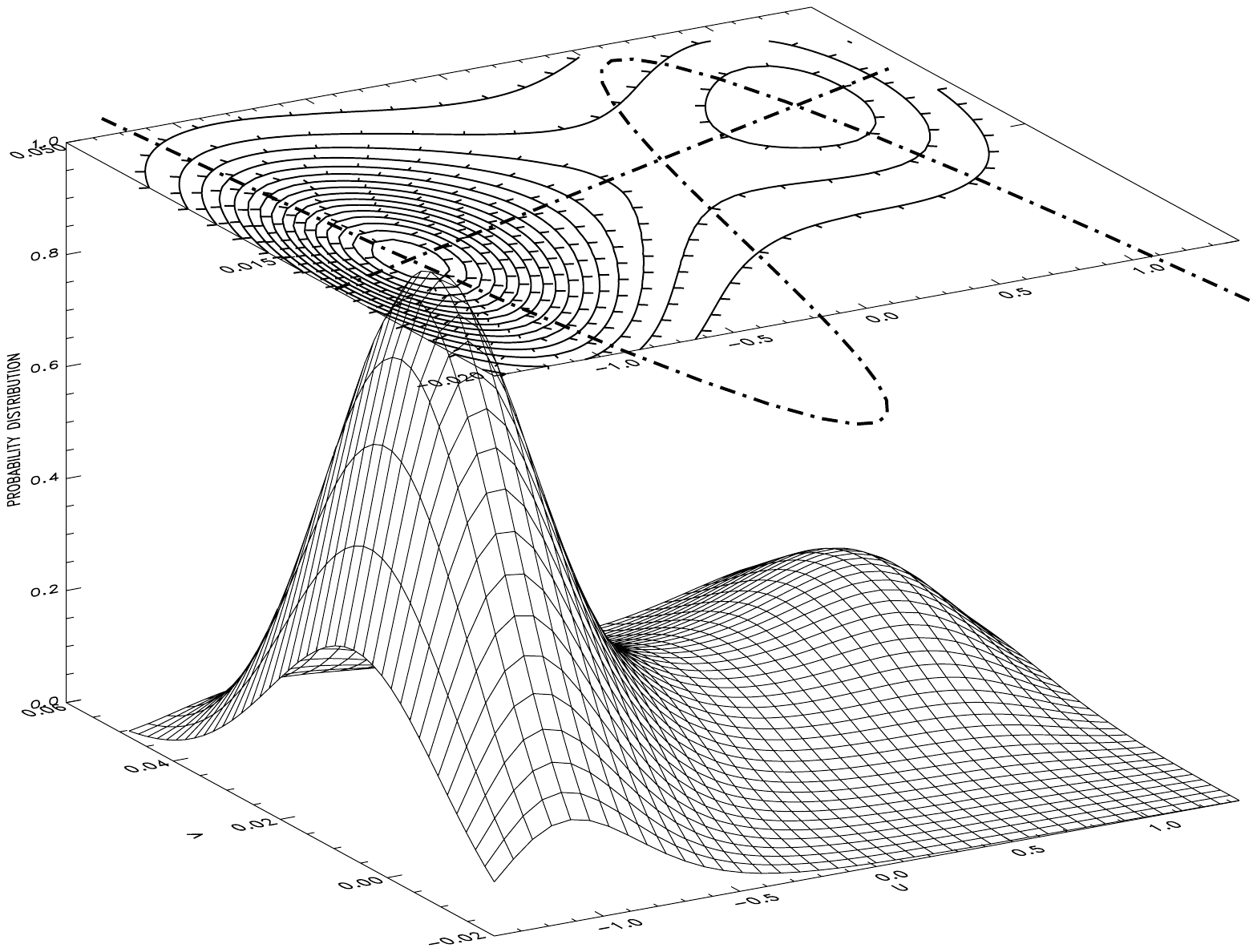}
  \caption{a) NEP in bistable regime; b) Stationary pdf in bistable regime.}
\end{figure}
Excitable dynamics can be conceptually decomposed into two phases, a \emph{fluctuation-dominated} one and a \emph{deterministic} one. It would be highly desirable to find a Lyapunov function, since it greatly simplifies the dynamical analysis. However, the existence of non-variational (or conserving) components in the phase-space flow is a hint that the integrability conditions fail for the purely deterministic system. This apparently insurmountable drawback was partially solved two decades ago by Graham and collaborators (see references in \cite{izdw98}) who defined the NEP for Langevin-type dynamics as the zero-noise limit of the logarithm of the \emph{stationary} probability density function (pdf). The extra freedom in the choice of the transport matrix can render in some cases the problem integrable. That is precisely the case for the space-independent FitzHugh--Nagumo model in its bistable and excitable regimes \cite{izdw98,izdw99}. This approach can be generalized to extended systems and the NEP associated to Eq.(\ref{eq:FHN}) (in the adiabatic limit, i.e.\ for slow signal) is \cite{said06}
\begin{eqnarray}
\Phi(t)=\Phi\{t,u_i(t),v_i(t)\}=\sum_{i=1}^N&&\left\{\frac{\epsilon}{\lambda_2}(v_i^2-2\beta\,u_iv_i-2Cv_i)+
\frac{2\lambda\epsilon}{\lambda_1\lambda_2}(\beta\,u_i^2+2Cu_i)\right.\nonumber\\
&&\left.-\frac{2}{\lambda_1}\left[\frac{a_c}{2}u_i^2-\frac{a_c}{4}u_i^4+S_g(t)\,u_i\right]+
2D\frac{u_iu_{i+1}}{\lambda_1}\right\},\label{eq:Phi}
\end{eqnarray}
which must obey the integrability condition $\beta\lambda_1+\lambda_2/\epsilon=2\lambda$ \cite{izdw98}.

Figure \ref{fig:2}b depicts (in full line) the time evolution of the NEP during the phase-space excursion starting at the upper initial condition in Fig.\ \ref{fig:2}a, together with that of $u$ (dashed line) and $v$ (dotted line). We remark that in Figs.\ \ref{fig:2}a and \ref{fig:2}b there is no noise and $\Phi(t)$ is the Lyapunov functional of the deterministic dynamics. Figures \ref{fig:3}a and \ref{fig:3}b (respectively \ref{fig:4}a and \ref{fig:4}b) are 3D and contour plots of the NEP and the corresponding stationary pdf for the excitable (respectively bistable) regime.
\section{\label{sec:3}Results for the coupled system}
\begin{figure}\label{fig:5}
  \includegraphics[height=.3\textheight,bb= 54pt 360pt 558pt 720pt]{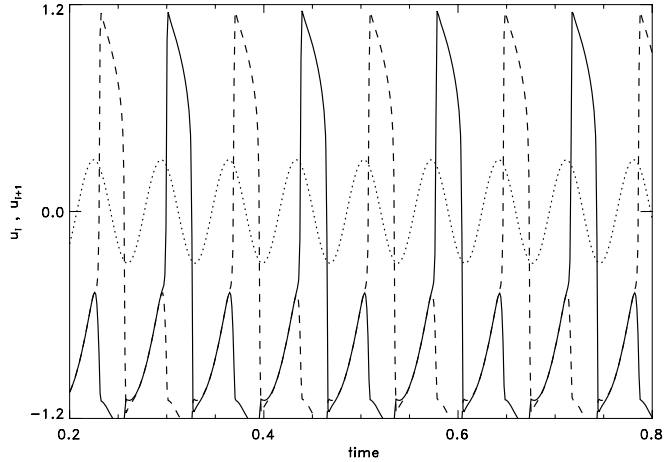}
  \caption{Time evolution of $u$ for two neighbor neurons.}
\end{figure}
Synchronization between the coupled system and the external signal is observed above some noise-intensity threshold. Figure \ref{fig:5} is a plot of the time evolution of $u_i$ (full line), together with that of $u_{i+1}$ (dashed line) for a given neuron $i$, showing their phase relation to the signal (dotted line)\footnote{The signal has been augmented in about two orders of magnitude and shifted to aid the sight.} According to Fig.\ \ref{fig:2}, we may call ``active'' those cells $i$ for which $u_i(t)$ exceeds some threshold value $u_{\mathrm{th}}$. Because of the coupling, as one neuron becomes active, it inhibits the activation of its nearest neighbors. The perfect alternance seen in the figure may fail because of the noise, a necessary ingredient for the activation.

\begin{figure}\label{fig:6}
  \includegraphics[height=.2\textheight,bb= 54pt 360pt 558pt 720pt]{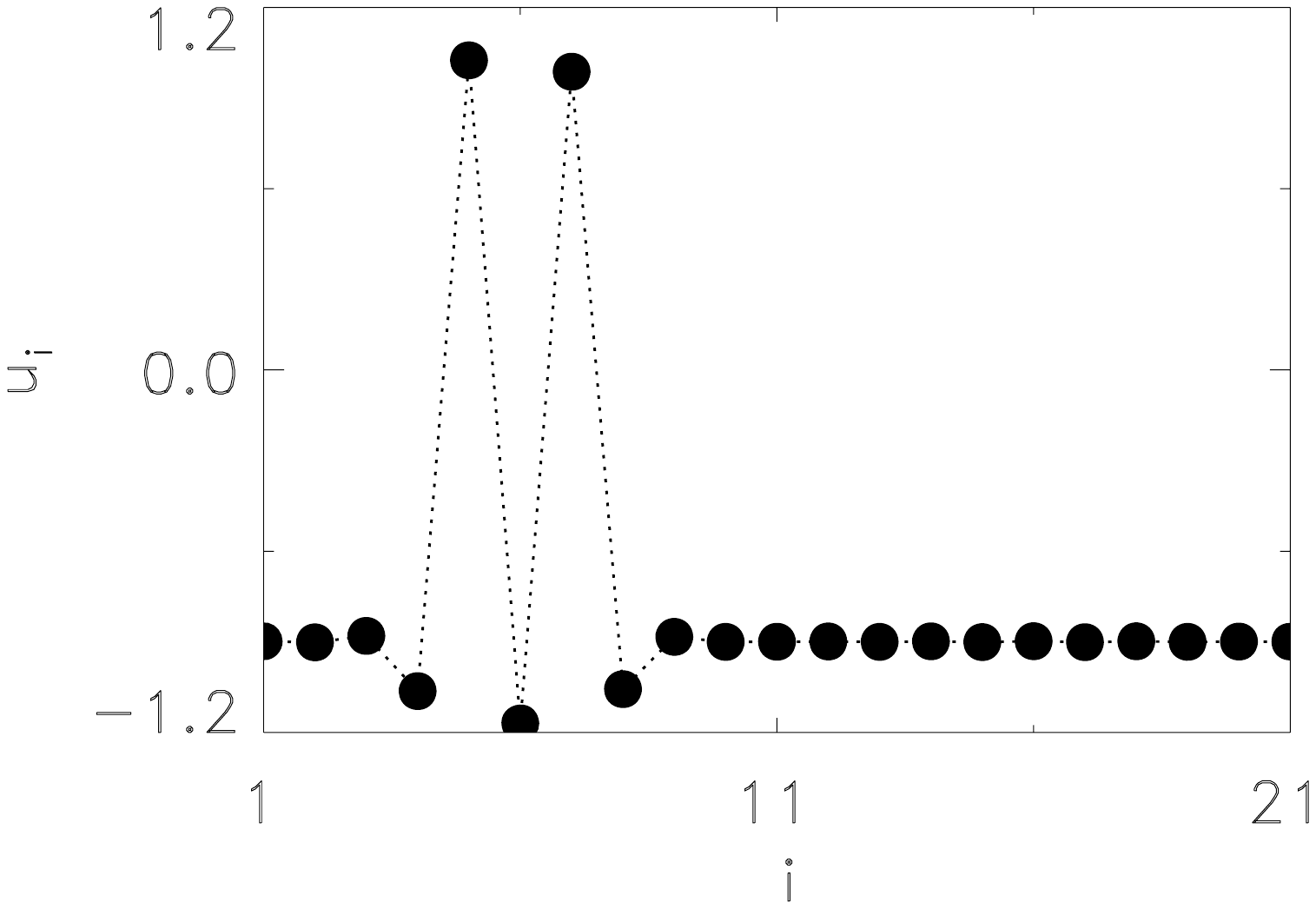}
  \includegraphics[height=.2\textheight,bb= 54pt 360pt 558pt 720pt]{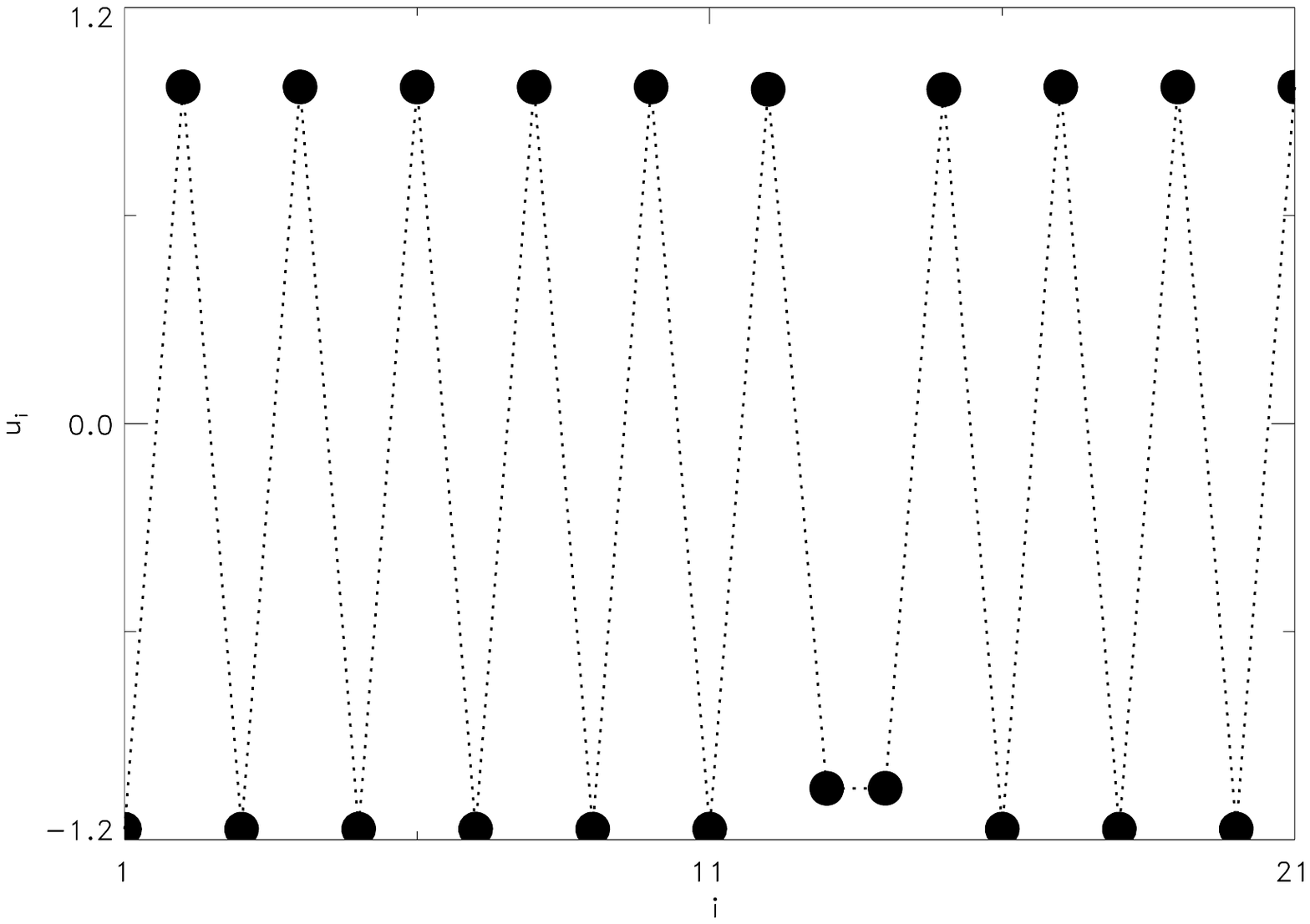}
  \caption{Two snapshots of $\{u_i\}$ showing different degrees of synchrony.}
\end{figure}

\begin{figure}\label{fig:7}
  \includegraphics[height=.2\textheight,bb= 54pt 360pt 558pt 503pt]{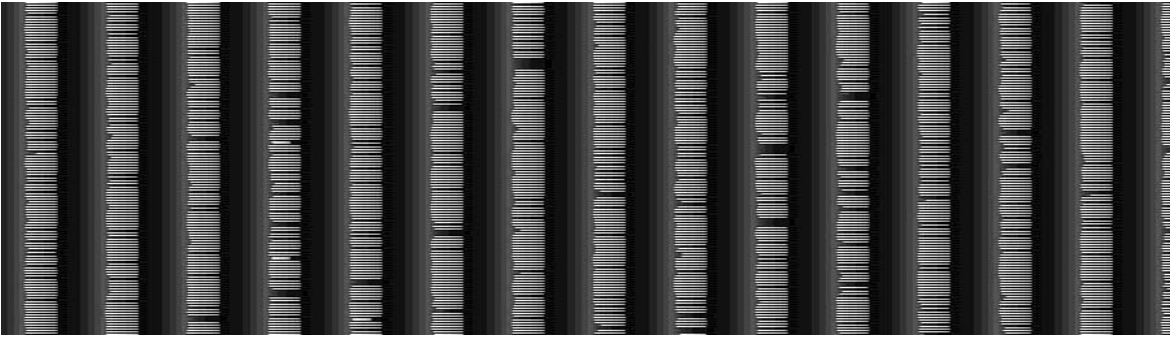}
  \caption{Synchronization of the ring. White corresponds to activation and black to inhibition. Horizontal dimension corresponds to time and vertical to space.}
\end{figure}

A detail of the alternance can be seen in Fig.\ \ref{fig:6} for an $N'=21$ subset of the ring. Figure \ref{fig:6}a shows a situation (snapshot) of poor synchronization, in which only two neurons are active; Fig.\ \ref{fig:6}b exhibits a case of a ``kink'' in the synchronized configuration, induced by the fact that noises are local. Note that the kinks break locally the observed coherence, and the complete history of the time evolution can be followed as a record of activity (see Fig.\ \ref{fig:7}).

\begin{figure}\label{fig:8}
  \includegraphics[height=.2\textheight,bb= 54pt 360pt 558pt 720pt]{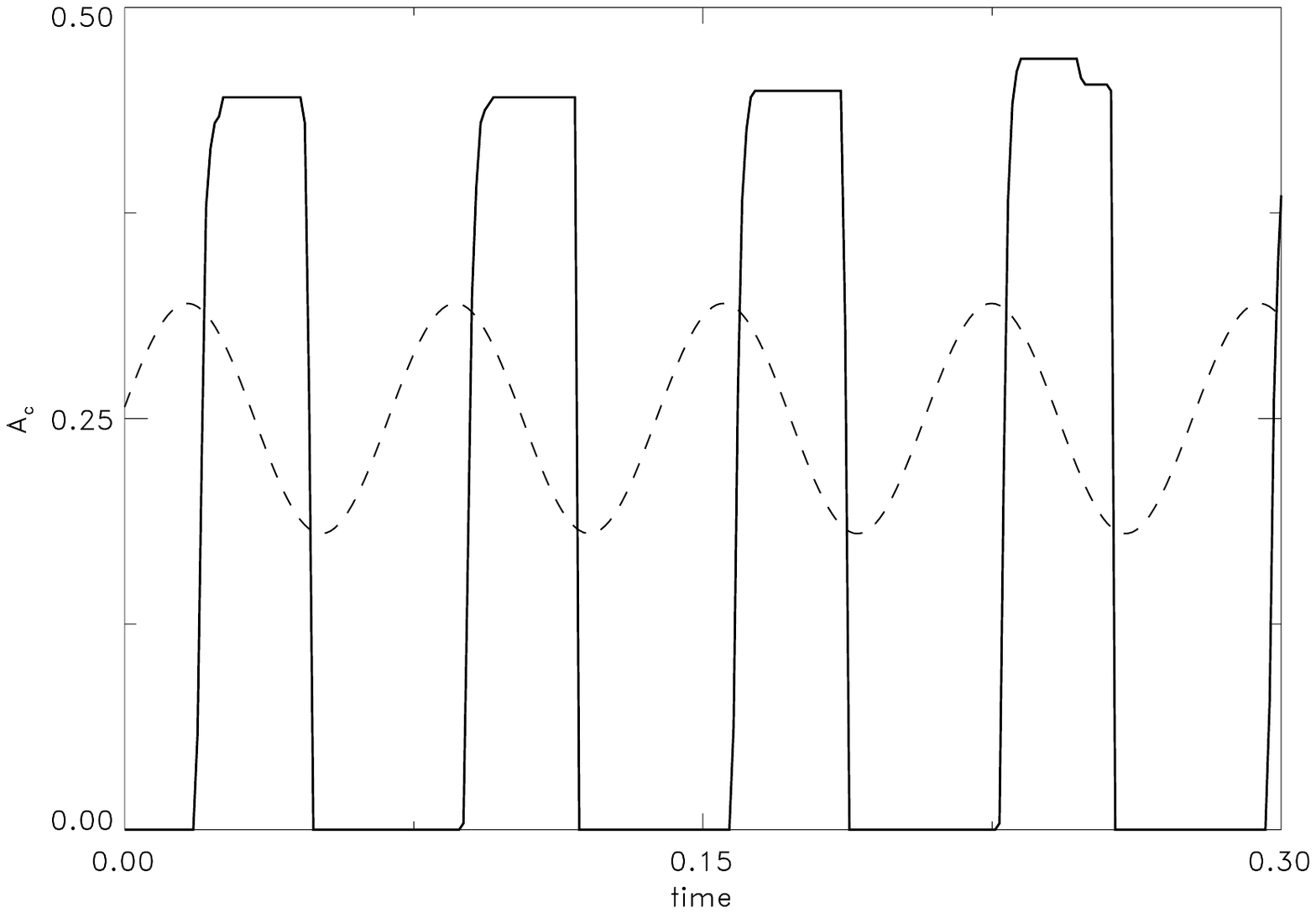}
  \includegraphics[height=.2\textheight,bb= 54pt 360pt 558pt 720pt]{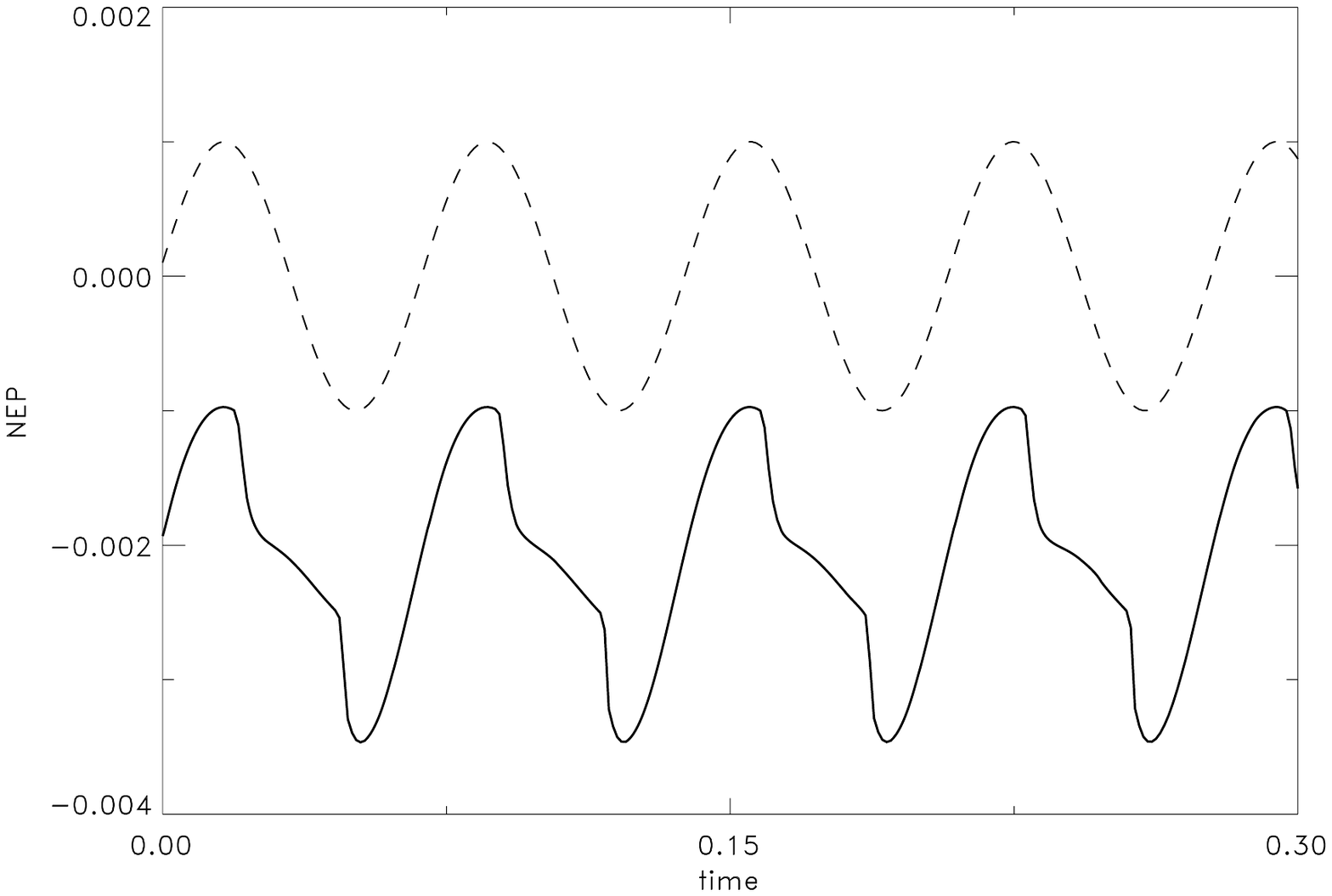}
  \caption{a) $Ac$ vs $t$ for high synchronization; b) Time evolution of the NEP}
\end{figure}

A measure of ``activity'' for the whole ring is
\begin{equation}\label{eq:act}
Ac(t)=\frac{1}{N}\sum_{i=1}^N\theta[u_i(t)-u_{\mathrm{th}}].
\end{equation}
In perfect synchrony, $Ac=0.5$. Note that since the signal is subthreshold for the coupled system, $Ac=0$ below threshold. Figure \ref{fig:8}a depicts the activity as a function of time for a fixed noise intensity, showing again its phase relationship with the signal (dashed line). In Fig.\ \ref{fig:8}b we show the NEP for the whole ring as a function of time, together with the (scaled) signal for reference. We remark that the observed dynamical symmetry breakdown decreases the Lyapunov function of the whole ring with respect to that of the homogeneous state, providing the route to stable synchronization.

\begin{figure}\label{fig:9}
  \includegraphics[height=.2\textheight,bb= 54pt 360pt 558pt 720pt]{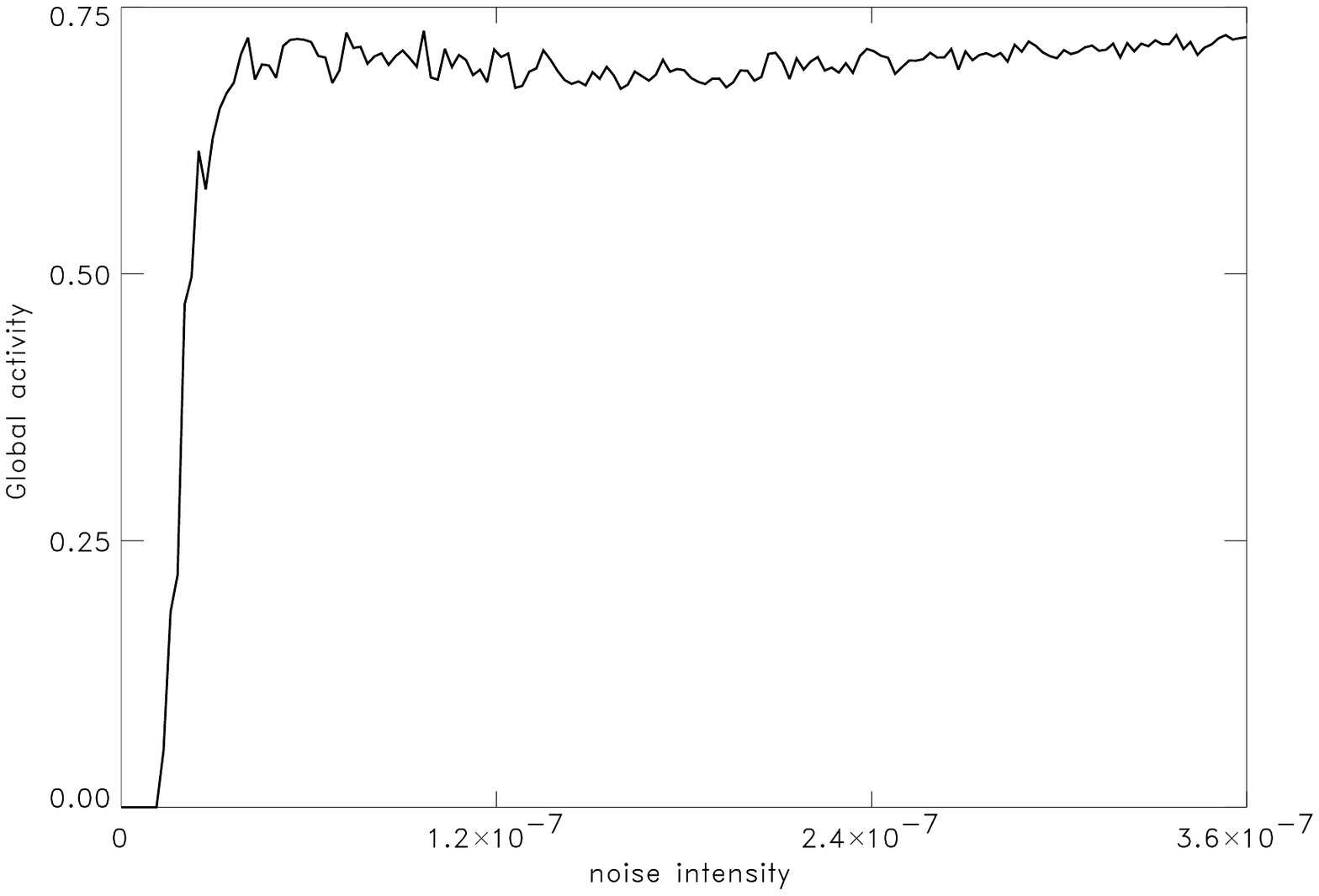}
  \includegraphics[height=.2\textheight,bb= 54pt 360pt 558pt 720pt]{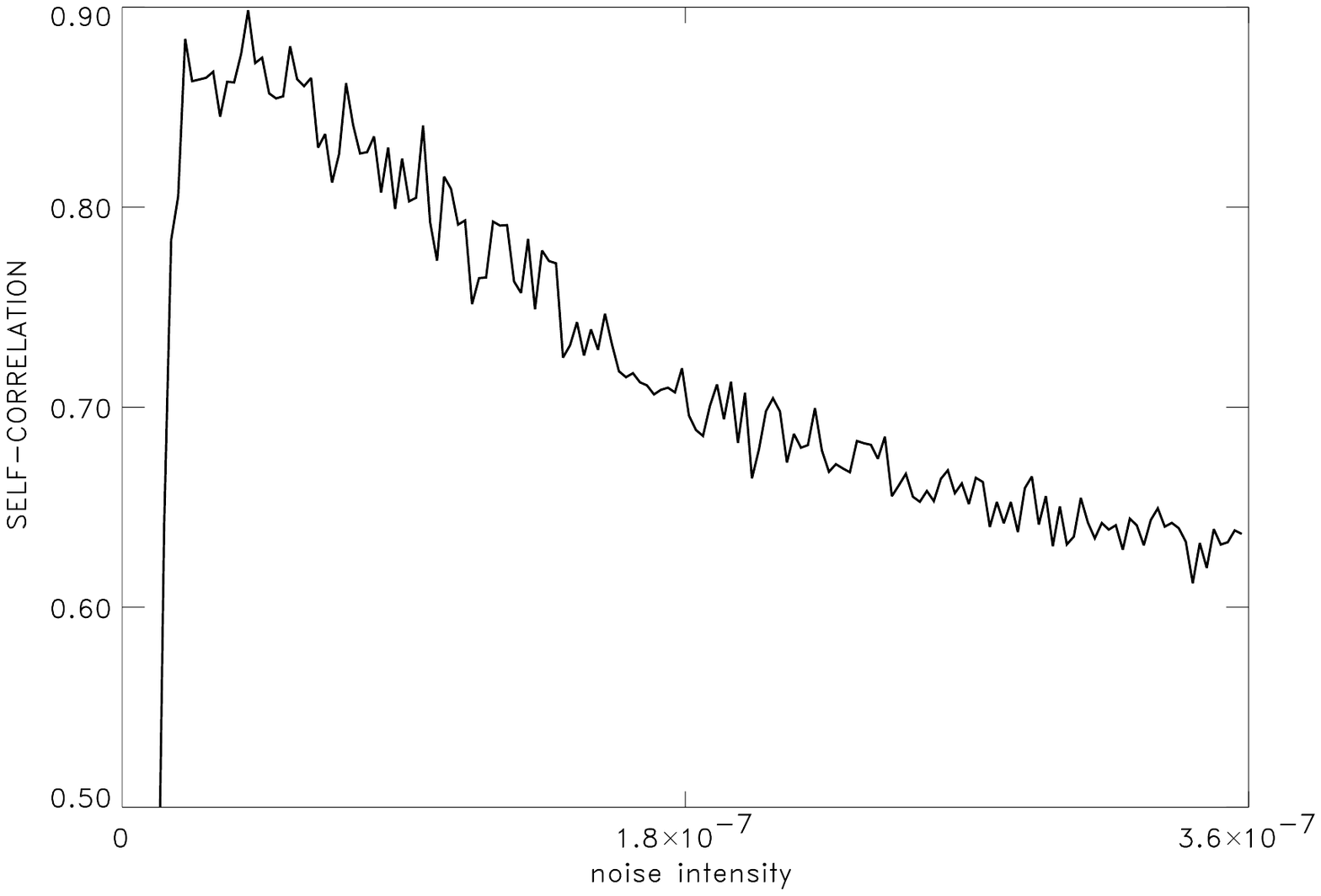}
  \caption{a) $G_a$ vs $\eta$; b) $C$ vs $\eta$.}
\end{figure}

A global estimator of synchronization can be defined as
\begin{equation}\label{eq:act}
G_a=\frac{\int_0^{t_f}Ac(t)dt}{0.5N\,t_f}.
\end{equation}
Figure \ref{fig:9}a is a plot of $G_a$ as a function of the noise intensity. The existence of a threshold value of noise intensity and of a saturation effect can be clearly seen. The noise intensities are low enough not to degrade the excitable dynamics.

Numerical simulations indicate that the coherence of firing decreases with the noise intensity although the global activity (representative of global estimators) keeps the order of magnitude. To quantify this phenomena we have calculated the normalized self-correlation $C=\langle u_iu_{i+2}\rangle$ as a function of the  noise intensity $\eta$. As we show in Fig. \ref{fig:9}b the system shows a kind of ``stochastic resonance in coherence'' that cannot be inferred from measures of global activity.
\section{\label{sec:4}Conclusions}
We have investigated the noise-induced synchronization with an external signal of a ring of phase-repulsively coupled FHN elements. We have derived the exact NEP of the extended system and the observed symmetry breakdown was related with the Lyapunov-functional properties of the NEP. We remark that the same conclusion holds qualitatively for the work in  Ref.\ \cite{bcnm01}. Although the observed phenomenon is noise-sustained and global activity increases with noise intensity, a degradation of coherence can be appreciated.
\begin{theacknowledgments}
Financial support from CONICET, ANPCyT and the National University of Mar del Plata is acknowledged.
\end{theacknowledgments}

\end{document}